%% file: main.tex
\documentclass[sigconf,nonacm]{acmart}
\setkeys{acmart.cls}{balance=false}

\usepackage{array}
\usepackage{float}
\usepackage{xurl}
\setcopyright{none}
\acmDOI{}
\acmISBN{}
\authorsaddresses{}

\hypersetup{
  hidelinks,
  pdftitle={DASWeave: Weaving Evidence across Speaking Conditions for Multimodal Depression, Anxiety, and Stress Screening},
  pdfauthor={Binyan Chen; Tong Pan; Yadong Liu; Xin Yun; Shangfei Wang}
}

\title[DASWeave]{DASWeave: Weaving Evidence across Speaking Conditions for Multimodal Depression, Anxiety, and Stress Screening}

\author{Binyan Chen}
\affiliation{%
  \institution{University of Science and Technology of China}
  \city{Hefei}
  \country{China}}
\email{sztn@mail.ustc.edu.cn}
\author{Tong Pan}
\affiliation{%
  \institution{University of Science and Technology of China}
  \city{Hefei}
  \country{China}}
\email{pantong@mail.ustc.edu.cn}
\author{Yadong Liu}
\affiliation{%
  \institution{University of Science and Technology of China}
  \city{Hefei}
  \country{China}}
\email{yadongliu@mail.ustc.edu.cn}
\author{Xin Yun}
\affiliation{%
  \institution{University of Science and Technology of China}
  \city{Hefei}
  \country{China}}
\email{yx9329@mail.ustc.edu.cn}
\author{Shangfei Wang}
\authornote{Corresponding author: sfwang@ustc.edu.cn.}
\affiliation{%
  \institution{University of Science and Technology of China}
  \city{Hefei}
  \country{China}}
\email{sfwang@ustc.edu.cn}

\begin{document}

\begin{abstract}
\input{sections/abstract}
\end{abstract}
\maketitle
\raggedbottom

\input{sections/introduction}
\input{sections/system}
\input{sections/challenge}
\input{sections/training}
\input{sections/results}
\input{sections/conclusion}

\clearpage
\bibliographystyle{ACM-Reference-Format-Unsrt}
\bibliography{references}

\clearpage
\appendix
\input{sections/appendix}

\end{document}

%% file: sections/abstract.tex
Multimodal screening for depression, anxiety, and stress in children and
adolescents uses verbal and nonverbal behaviour. Most existing multimodal
methods learn from a single recording or interview. Yet a participant's
behaviour changes with the speaking condition, and these changes hold
evidence that is invisible in a single recording. We propose DASWeave to
bring this evidence into a neural network that encodes each response and
then combines all responses of a participant. We train supervised trees to
find which behavioural statistics matter for screening, both within each
response and in how they change across conditions. We feed the tree
outputs into the network both before and after this combination. We
fine-tune a large language model to relate what a participant says to how
they say it across conditions, and feed its representation into the
network after the combination. The language model reads behaviour as text
descriptions. We therefore derive additional training targets for the
network from a multimodal large language model that takes in the
audio-visual features directly.
On the official hidden test set of the ACM Multimedia 2026 AdoDAS Grand
Challenge (Track~1), DASWeave ranks second among 95 eligible teams. Its mean
F1 is 27.1\% higher than that of the official audio-visual baseline. It
also has the highest mean AUROC among the top five teams. Ablation studies
on the validation set show that the supervised trees and the fine-tuned
language model contribute most to mean F1.

%% file: sections/introduction.tex
\section{Introduction}
\label{sec:introduction}

Depression and anxiety can disrupt school attendance and schoolwork, and
unaddressed mental health conditions in adolescence can have consequences
in adulthood. However, many such conditions remain unrecognized and
untreated~\cite{who2025adolescent}. Screening for depression, anxiety, and
stress (D/A/S) in children and adolescents can help identify those who
need support. Automatic screening can assist this process by analysing
what participants say and how they sound and look while speaking.

Research on automatic screening has mainly studied depression. Multimodal
depression detection estimates a person's depression status or severity
from a recording of them speaking. It combines two or more modalities,
such as the transcript, audio, and video. Existing methods follow two main
directions. The first designs network architectures for the task. These
networks can learn temporal patterns and cross-modal interactions related
to depression within the recording, such as the question--answer sequence
of an interview~\cite{interview_sequence,daic} and the interplay of audio
and visual cues over time~\cite{depmstat,mpcf}. However, they learn these
patterns only from labelled depression data, which are limited in size.
The second transfers knowledge from speech and language models pretrained
on large-scale data. Such models bring general knowledge of speech and
language to the task~\cite{speech_text_foundations}, and a large language
model (LLM) can further use psychological
knowledge~\cite{llm_spoken_depression}. Yet the pretraining of these
models serves general purposes rather than depression detection.

A participant's speech and facial behaviour change with the speaking
condition, such as reading a fixed text or answering an open question.
Such differences offer evidence that no single response contains. When
reading a fixed text, all participants say the same words, so the reading
reflects their habitual voice and expression. Differences between the open
responses and the reading are therefore less affected by individual
speaking habits. Comparing open responses with one another, such as
answers about happy and sad memories, shows how strongly a participant's
behaviour reacts to emotional content. Existing methods model behaviour
within each response and what participants say. Yet they have paid little
attention to behavioural differences across speaking conditions.

To address this gap, we propose DASWeave to bring behavioural differences
across speaking conditions into multimodal D/A/S screening. Its neural
networks and supervised trees learn patterns specific to D/A/S, while its
language models add knowledge from large-scale pretraining and are
fine-tuned for the task. Figure~\ref{fig:system} shows its structure. We
refer to each recorded response as a session. The \emph{Shared Session
Encoder} encodes each session from its audio-visual features. For open
answers, it adds the question and the transcript as text input. The
session representations are then aggregated into a participant
representation. \emph{Two-Level Tree Fusion} uses gradient-boosted
trees~\cite{lightgbm} to learn from thousands of behavioural statistics.
The tree outputs are added to each session representation and to the
participant representation. The participant-level statistics include
differences across speaking conditions. In \emph{Participant Language
Fusion}, a fine-tuned LLM reads all of a participant's answers. Its input
also describes how their speaking behaviour differs from the reading. The
resulting language representation is added to the participant
representation. During training, \emph{MLLM Auxiliary Supervision} gives
the main predictor additional targets derived from a multimodal LLM
(MLLM), which encodes audio-visual features and transcripts together.

We evaluate DASWeave on Track~1 of the ACM Multimedia 2026 AdoDAS Grand
Challenge~\cite{adodas2026}. It ranked second among 95 eligible teams on
the official hidden test set, with mean F1 of 0.5853 against the official
audio-visual baseline's 0.4604, and its mean AUROC of 0.8019 was the
highest among the top five teams. On the validation set, the full method
outperforms the Shared Session Encoder alone. Each of the other three
components improves F1 on all three targets, and Participant Language
Fusion and Two-Level Tree Fusion contribute the most.

Our contributions are as follows.
\begin{itemize}
  \item We make behavioural differences across speaking conditions an
    explicit source of evidence for multimodal D/A/S screening.
  \item We propose DASWeave, which uses supervised trees on behavioural
    statistics and a fine-tuned LLM on text descriptions to bring these
    differences into a neural predictor that encodes each response.
  \item DASWeave ranks second among 95 teams in AdoDAS Track~1, with a mean
    F1 27.1\% higher than the official baseline.
\end{itemize}

%% file: sections/system.tex
\section{Method}
\label{sec:system}

We study participant-level D/A/S screening from several speaking sessions.
Each participant contributes one session per speaking condition. The input
contains the audio-visual features of every session, pooled eGeMAPS
descriptors~\cite{egemaps}, and transcripts of the open responses. For
participant $i$, the model outputs three probabilities corresponding to the
binary labels $\mathbf{y}_i=(y_{i,D},y_{i,A},y_{i,S})\in\{0,1\}^3$.
In the AdoDAS evaluation, the four speaking conditions are reading (A01),
recounting the previous day (B01), and recalling the happiest (B02) and
saddest (B03) memories from the past week~\cite{adodas2026}.

\begin{figure*}[t]
  \centering
  \includegraphics[width=\textwidth]{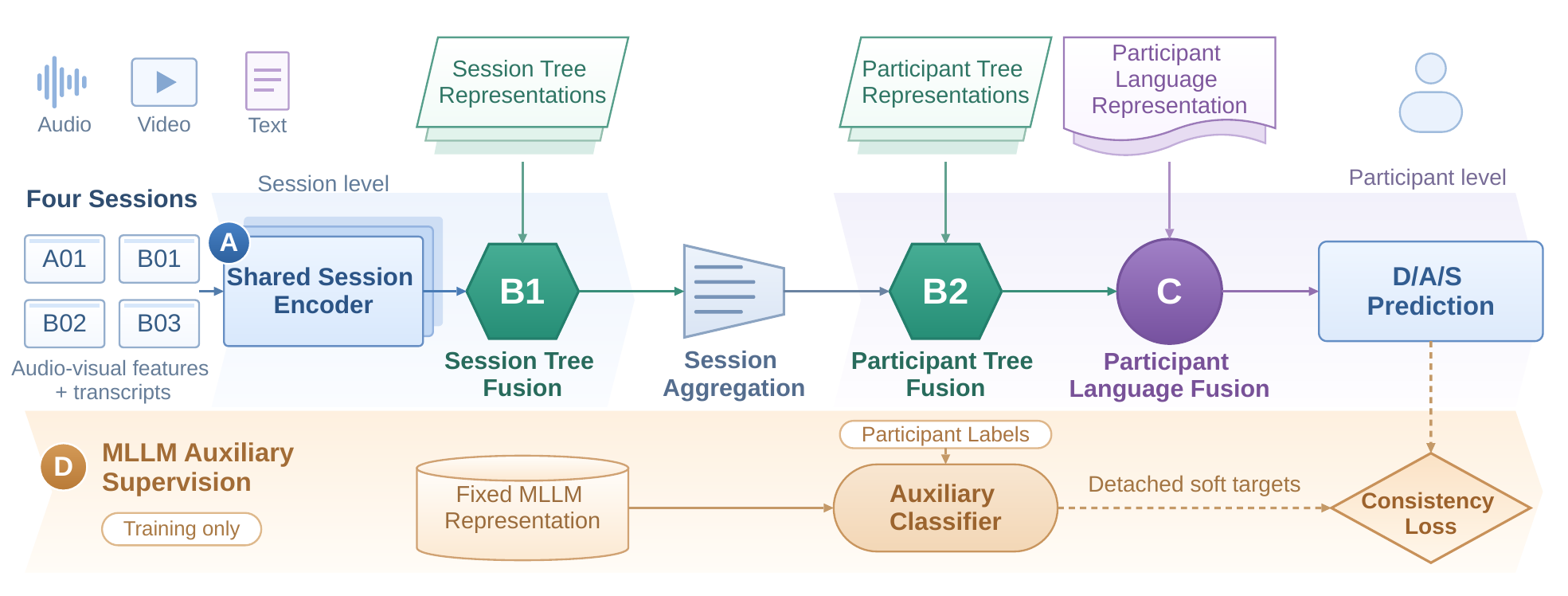}
  \Description{Four sessions pass through a shared audio-visual-text encoder,
    session tree fusion, session aggregation, participant tree fusion, and
    participant language fusion to produce D/A/S predictions. Three
    side-representation cards connect to their respective fusion modules.
    In the training-only MLLM auxiliary supervision group, a classifier
    learns from participant labels on a fixed representation and supplies
    detached soft targets to a consistency loss, which also receives the
    main predictions.}
  \caption{Overview of DASWeave for participant-level D/A/S screening.
    (A) Shared Session Encoder encodes behavioural dynamics and local text
    features of each session. (B1--B2) Two-Level Tree Fusion adds tree tokens
    of session and participant statistics before and after session
    aggregation. (C) Participant Language Fusion encodes all answers
    together with descriptions of how speaking behaviour differs from the
    reading. (D) MLLM Auxiliary Supervision
    trains an auxiliary classifier with participant labels on fixed MLLM
    representations. Its detached probabilities provide soft targets for the
    main predictor during training.
    Solid arrows denote representation flow; dashed arrows denote auxiliary
    supervision. D/A/S denote Depression, Anxiety, and Stress.}
  \label{fig:system}
\end{figure*}

\subsection{Method Overview}
\label{sec:overview}
Figure~\ref{fig:system} shows the four components of DASWeave and where
differences across speaking conditions enter the model. The \emph{Shared
Session Encoder} (A) encodes behavioural dynamics within each session,
together with local text features of open responses, in a session
representation $\mathbf{h}_s$. \emph{Two-Level Tree Fusion} uses
supervised trees to encode behavioural statistics. Session Tree Fusion
(B1) adds tree tokens of session statistics before masked session
aggregation. After aggregation, Participant Tree Fusion (B2) adds tree
tokens of participant statistics, which include differences across
speaking conditions.

\emph{Participant Language Fusion} (C) uses a fine-tuned language model to
encode all answers together with text descriptions of speaking behaviour,
including differences from the reading baseline. A learned residual update
adds the resulting language representation to the participant
representation for D/A/S prediction. Components A--C form the prediction
path. During training, \emph{MLLM Auxiliary Supervision} (D) brings joint
audio-visual-text information into the main model through an auxiliary
classifier on fixed MLLM representations. The classifier is trained with
participant labels, and its detached probabilities serve as soft targets
for the main prediction head.

We fit the tree, language, and MLLM side branches before main training.
These branches learn from the same D/A/S labels as the main model. Each
training participant therefore receives out-of-fold (OOF) side inputs from
models fitted on the other folds, which have not seen that participant's
labels. With these inputs fixed, we jointly train the session encoder, leaf
embeddings, fusion modules, and prediction heads.
Section~\ref{sec:side_training} gives the fitting and extraction protocol.

\subsection{Shared Session Encoder}
\label{sec:backbone}
The Shared Session Encoder (A) maps each session to a 256-dimensional
representation $\mathbf{h}_s$. It encodes behavioural dynamics from the
audio-visual features. For open responses, it adds the question and
transcript as local text features. Sharing the encoder across the four
speaking conditions places all session representations in a common space
for aggregation, and a session-type embedding identifies each condition.
We encode audio and video separately through feature adaptation, temporal
convolution, and attentive statistics pooling, and then fuse the two
summaries with the local text features.

\textbf{Feature adaptation and temporal encoding.}
The audio stream uses log-Mel descriptors, mel-frequency cepstral
coefficients (MFCCs), voice activity detection (VAD) features, and Chinese XLSR
wav2vec~2.0 embeddings~\cite{wav2vec2,xlsr}. For video, the inputs are head
pose, facial behaviour, quality indicators, aggregated voice activity,
and SigLIP-SO400M embeddings~\cite{siglip}. All sequence features align to
a common 25~Hz timeline, with separate validity masks for audio and video.

Each feature group has an adapter consisting of layer normalization, a
linear projection, GELU, and dropout, which maps the sequence to 64 channels.
We concatenate the adapted groups within each modality and project them to
256 channels before applying separate audio and video temporal
convolutional networks (TCNs)~\cite{tcn}. Each TCN has six residual blocks,
with two convolutions per block, kernel size three, and dilation factors
$1,2,4,8,16,32$. The increasing dilation expands the temporal context of
successive blocks. We apply validity masks before temporal encoding and
after every residual block.

\textbf{Attentive statistics pooling.}
An attentive mean and standard deviation~\cite{attentive_pooling}
summarize each encoded sequence, retaining both its typical feature level
and the variation within a response. For modality
$m\in\{\mathrm{a},\mathrm{v}\}$, let $\mathbf{f}^{(m)}_{s,t}$ denote the
encoded frame at time $t$ and $\mathcal{V}^{(m)}_s$ the valid frame indices.
The pooling weights are
\begin{equation}
  a^{(m)}_{s,t} =
  \operatorname{softmax}_{t\in\mathcal{V}^{(m)}_s}
  \left(\mathbf{w}_m^{\top}\mathbf{f}^{(m)}_{s,t}
       +\alpha_m v_{s,t}+\beta_m q_{s,t}\right),
  \label{eq:pooling_weights}
\end{equation}
where $v_{s,t}$ and $q_{s,t}$ are the aligned voice activity and video quality
signals. We learn separate projections $\mathbf{w}_m$ and coefficients
$\alpha_m,\beta_m$ for the two modalities. Voice activity and quality
affect frame weighting. The validity mask determines which frames
participate in pooling. This gives the representation
\begin{equation}
  \begin{aligned}
    \boldsymbol{\mu}^{(m)}_s & =
      \sum_{t\in\mathcal{V}^{(m)}_s}
      a^{(m)}_{s,t}\mathbf{f}^{(m)}_{s,t},\\
    \boldsymbol{\nu}^{(m)}_s & =
      \sum_{t\in\mathcal{V}^{(m)}_s} a^{(m)}_{s,t}
      \big(\mathbf{f}^{(m)}_{s,t}-\boldsymbol{\mu}^{(m)}_s\big)^{\odot2},\\
    \mathbf{g}^{(m)}_s & =
      \big[\boldsymbol{\mu}^{(m)}_s;
      \sqrt{\max(\boldsymbol{\nu}^{(m)}_s,\epsilon)}\big].
  \end{aligned}
  \label{eq:attentive_statistics}
\end{equation}
The square, maximum, and square root are element-wise operations,
$\epsilon=10^{-8}$, and each modality summary $\mathbf{g}^{(m)}_s$ has 512
dimensions.

\textbf{Local text encoding and session fusion.}
A frozen Chinese MacBERT-base encoder~\cite{macbert} reads the question and
the automatic speech recognition (ASR) transcript of each open response as
a sentence pair. Layer normalization, a linear layer, GELU, and dropout
form a trainable projection from the 768-dimensional \texttt{[CLS]} state
to a 64-dimensional text vector $\mathbf{c}_s$. For reading,
$\mathbf{c}_s=\mathbf{0}$.
A learned adapter maps the pooled eGeMAPS descriptors to a 64-dimensional
acoustic vector $\mathbf{p}_s$, and a learned 16-dimensional session-type
embedding $\mathbf{e}_s$ identifies the speaking condition. A multilayer
perceptron (MLP) with two linear layers combines these inputs and the audio
and video summaries:
\begin{equation}
  \mathbf{h}_s = \operatorname{MLP}_{\mathrm{sess}}
    \big([\mathbf{g}^{(\mathrm{a})}_s;
      \mathbf{g}^{(\mathrm{v})}_s;\mathbf{p}_s;
      \mathbf{c}_s;\mathbf{e}_s]\big).
  \label{eq:session_encoding}
\end{equation}

\subsection{Two-Level Tree Fusion}
\label{sec:tree}
Two-Level Tree Fusion (B1--B2) uses supervised trees to encode behavioural
statistics at the session and participant levels. Session statistics
summarize feature levels and variability within a response. Participant
statistics also describe variation and differences across speaking
conditions. At each level, separate LightGBM classifiers~\cite{lightgbm}
are fitted to these statistics for D/A/S prediction. Each tree splits on
statistic thresholds chosen under D/A/S supervision, so the leaf that a
sample reaches records a label-relevant combination of statistics. We
represent each sample by its leaf indices~\cite{tree_feature_encoding} and
fuse learned embeddings of these indices with the corresponding session or
participant representation.

\textbf{Tree inputs at two levels.}
For the session branch, we pool HuBERT-large audio
embeddings~\cite{hubert} and ViT-base visual embeddings~\cite{vit} into
means and standard deviations. These enter alongside eGeMAPS, log-Mel/MFCC
statistics, voice activity, head pose, facial behaviour, and quality
descriptors. Four session-type indicators complete the 4,207 features per
session. Each session uses its participant's D/A/S labels for supervision.

The participant branch pools HuBERT-large and DINOv2-large
embeddings~\cite{dinov2} in the same way and uses the same low-dimensional
audio-visual statistics as the session branch. We concatenate the four
session feature vectors with their element-wise mean, standard deviation,
minimum, and maximum, and with two differences between conditions. The
difference $\mathrm{B01}-\mathrm{A01}$ compares daily narration with the
reading baseline, and $\mathrm{B03}-\mathrm{B02}$ compares sad with happy
recall. With 47,150 features, the participant classifiers can use both
individual conditions and differences across speaking conditions.
Section~\ref{sec:side_training} gives the tree fitting settings.

\textbf{Leaf embeddings and tree tokens.}
Trainable leaf embeddings map the supervised tree partitions to dense
vectors. DeepGBM~\cite{deepgbm} also uses dense leaf representations as the
basis for tree-to-network distillation. Here, we learn the embeddings
jointly with the main D/A/S predictor while the tree classifiers stay
fixed, producing one tree token per target at each level. Separate leaf
encoders serve the session and
participant branches. Each encoder assigns a 32-dimensional embedding
table to every target and tree position, sharing these tables across folds.
For a sample and target $\ell\in\{D,A,S\}$, let $\lambda_{\ell,f,k}$ be
the leaf index reached in tree $k$ of fold $f$, and $\mathcal{F}_\ell$ the
folds available for that sample and target. We average the embedding vectors
over trees and available folds:
\begin{equation}
  \mathbf{e}_{\ell} =
  \frac{1}{\max(1,|\mathcal{F}_\ell|)}
  \sum_{f\in\mathcal{F}_\ell}\frac{1}{K}
  \sum_{k=1}^{K}\mathbf{E}_{\ell,k}
  [\lambda_{\ell,f,k}],
  \label{eq:leaf_encoding}
\end{equation}
where $K$ is the branch's tree count. We project $\mathbf{e}_\ell$ to a
256-dimensional tree token $\mathbf{t}_\ell$ with a linear layer, dropout,
and layer normalization. For a training participant, $\mathcal{F}_\ell$
contains only the fold whose model did not see that participant. For
validation and test participants, it contains all five folds.

\textbf{Session Tree Fusion (B1) and aggregation.}
Before aggregation, Session Tree Fusion combines the tree tokens of session
statistics with $\mathbf{h}_s$. For session $s$, the three tree tokens form
$\mathbf{T}^{\mathrm{S}}_s\in\mathbb{R}^{3\times256}$. Four-head
attention~\cite{transformer} uses the session representation
$\mathbf{h}_s$ as its query and the tree tokens as keys and values:
\begin{equation}
  \widetilde{\mathbf{h}}_s = \mathbf{h}_s +
  \boldsymbol{\gamma}_{\mathrm{S}}\odot
  \operatorname{MHA}\left(
    \operatorname{LN}(\mathbf{h}_s),
    \operatorname{LN}(\mathbf{T}^{\mathrm{S}}_s),
    \operatorname{LN}(\mathbf{T}^{\mathrm{S}}_s)\right).
  \label{eq:session_tree_fusion}
\end{equation}
The session query determines how the three tree tokens are combined.
We add the scaled attention output to $\mathbf{h}_s$ and aggregate the
resulting session representations. Let
$\mathcal{S}=\{\mathrm{A01},\mathrm{B01},\mathrm{B02},\mathrm{B03}\}$
and let $m_s\in\{0,1\}$ indicate session validity. We take a masked mean
and apply a two-layer MLP to obtain the participant representation:
\begin{equation}
  \mathbf{u}_0 = \operatorname{MLP}_{\mathrm{agg}}
  \left(\frac{\sum_{s\in\mathcal{S}}m_s\widetilde{\mathbf{h}}_s}
  {\max(1,\sum_{s\in\mathcal{S}}m_s)}\right).
  \label{eq:participant_aggregation}
\end{equation}

\textbf{Participant Tree Fusion (B2).}
After aggregation, Participant Tree Fusion adds the tree tokens of
participant statistics, which include differences across speaking
conditions. The
participant-level leaf encoder produces the tokens
$\mathbf{t}^{\mathrm{P}}_D$, $\mathbf{t}^{\mathrm{P}}_A$, and
$\mathbf{t}^{\mathrm{P}}_S$. We concatenate them with $\mathbf{u}_0$ and
pass the result through a two-layer MLP for a second residual update:
\begin{equation}
  \mathbf{u}_{\mathrm{T}} = \mathbf{u}_0 +
  \boldsymbol{\gamma}_{\mathrm{P}}\odot
  \operatorname{MLP}_{\mathrm{P}}
  \left([\mathbf{u}_0;\mathbf{t}^{\mathrm{P}}_D;
    \mathbf{t}^{\mathrm{P}}_A;\mathbf{t}^{\mathrm{P}}_S]\right).
  \label{eq:participant_tree_fusion}
\end{equation}
We initialize the learned channel-wise scales
$\boldsymbol{\gamma}_{\mathrm{S}}$ and $\boldsymbol{\gamma}_{\mathrm{P}}$
to 0.005, so each fused branch starts as a small residual whose
contribution is then learned. The output
$\mathbf{u}_{\mathrm{T}}\in\mathbb{R}^{256}$ passes to Participant Language
Fusion (C).

\subsection{Participant Language Fusion}
\label{sec:language}
Participant Language Fusion (C) uses a fine-tuned Qwen3-8B
model~\cite{qwen3} to encode all answers of a participant together with
text descriptions of their speaking behaviour. These descriptions include
how the participant's behaviour differs from the reading, so the model can
relate what is said to how it is said across conditions.
Before D/A/S prediction, a learned projection and residual update add the
resulting language representation to $\mathbf{u}_{\mathrm{T}}$.

\textbf{Participant prompt.}
The Chinese prompt follows session order. Acoustic and visual summaries
of A01 serve as the reading baseline. For each open response, the prompt
includes the question, ASR transcript, SenseVoice emotion
tags~\cite{sensevoice}, and acoustic and visual behavioural summaries.
Cross-session summaries cover response length, short-response counts, and
ratios of loudness and pitch range in open responses to the same
participant's reading. Training-derived quantile tags indicate where
numerical measurements fall within the training distribution. Task
instructions direct the model toward D/A/S cues. The prompt ends with a
fixed Chinese summary phrase, so the representation is always read at the
same position. Appendix~\ref{sec:appendix_prompts} describes the prompt
structure in English.

\textbf{Fine-tuning and representation extraction.}
We fine-tune Qwen3 with low-rank adaptation (LoRA)~\cite{lora} and a
three-output D/A/S classification head. The head computes
D/A/S logits from $\mathbf{e}_{\mathrm{L}}$, the 4,096-dimensional hidden
state at the final prompt token. The OOF states stay fixed during main
training. For validation and test, states come from a model fitted on the
complete official training split, following the extraction protocol in
Section~\ref{sec:side_training}.

\textbf{Residual language fusion.}
We apply layer normalization and a two-layer MLP with 512 hidden units,
GELU, and dropout to project $\mathbf{e}_{\mathrm{L}}$ to a 256-dimensional
token $\mathbf{z}_{\mathrm{L}}$. Four-head attention uses this token as its
single key/value input. We retain the output corresponding to the
participant query $\mathbf{u}_{\mathrm{T}}$:
\begin{equation}
  \begin{aligned}
    \mathbf{z}_{\mathrm{L}} &=
      \operatorname{MLP}_{\mathrm{L}}
      \big(\operatorname{LN}_{\mathrm{L}}(\mathbf{e}_{\mathrm{L}})\big),\\
    \mathbf{r}_{\mathrm{L}} &=
      \operatorname{MHA}\big(
        \operatorname{LN}_{q}(\mathbf{u}_{\mathrm{T}}),
        \operatorname{LN}_{kv}(\mathbf{z}_{\mathrm{L}}),
        \operatorname{LN}_{kv}(\mathbf{z}_{\mathrm{L}})\big).
  \end{aligned}
  \label{eq:language_residual}
\end{equation}
During main training, we learn the projector, attention parameters, and
channel-wise residual scale $\boldsymbol{\gamma}_{\mathrm{L}}$, which starts
at 0.005. The fused participant representation is
\begin{equation}
  \mathbf{u}_{\star} = \mathbf{u}_{\mathrm{T}} +
  \boldsymbol{\gamma}_{\mathrm{L}}\odot\mathbf{r}_{\mathrm{L}}.
  \label{eq:language_fusion}
\end{equation}
Since attention has one key/value token, the block supplies a projected
language residual, and the learned residual scale controls its contribution
relative to the other representations.

\subsection{MLLM Auxiliary Supervision}
\label{sec:mllm}
MLLM Auxiliary Supervision (D) brings joint audio-visual-text information
into training. A separate fine-tuned Qwen3-8B model encodes audio and
visual feature tokens together with response text. During main training,
its participant representations stay fixed, and an auxiliary D/A/S
classifier learns from them with participant labels. The classifier's
detached probabilities serve as soft targets for the prediction path formed
by components A--C.

\textbf{Multimodal representation construction.}
The MLLM takes HuBERT-large audio features and DINOv2-base visual features
from all four speaking sessions. Two projectors convert the pooled audio
and visual descriptors to one token per modality and session, producing
eight tokens in total. The corresponding sequences pass through a
Q-Former resampler~\cite{blip2}, where two learned queries
attend to each modality--session sequence to produce sixteen more tokens.
Session and modality embeddings identify their sources. These 24
multimodal tokens precede the text embeddings of a Chinese prompt with
the questions, transcripts, and emotion tags of the three open responses.

While fitting the MLLM branch, we train its multimodal projectors, LoRA
adapter, and D/A/S classification head using participant labels. After
fitting, the 4,096-dimensional hidden state at the final text token gives
the participant representation
$\mathbf{e}_{\mathrm{M}}$. We cache OOF states following the fitting and
extraction procedure in Section~\ref{sec:side_training}.

\textbf{Auxiliary classification.}
We use the fine-tuned MLLM as a fixed source of representations and learn a
new prediction mapping on its OOF states under the main training
objective. During main model training, an auxiliary classifier maps the
normalized MLLM state to three D/A/S logits:
\begin{equation}
  \begin{aligned}
    \bar{\mathbf{e}}_{\mathrm{M}} &={}
      \mathbf{e}_{\mathrm{M}}/\|\mathbf{e}_{\mathrm{M}}\|_2,\\
    \mathbf{z}^{\mathrm{M}} &={}
      \mathbf{W}_{\mathrm{M}}g_{\mathrm{M}}
      (\bar{\mathbf{e}}_{\mathrm{M}})+\mathbf{b}_{\mathrm{M}}
      \in\mathbb{R}^{3}.
  \end{aligned}
  \label{eq:auxiliary_logits}
\end{equation}
Here, $g_{\mathrm{M}}$ applies layer normalization, a projection to 256
dimensions, GELU, and dropout, and a final linear layer gives the logits.
Participant labels supervise the classifier with the main head's
positive-class weights $w_\ell$, defined in Section~\ref{sec:objective}.
The MLLM's own head used the ratio $N^-_\ell/N^+_\ell$ of each training
fold as its positive-class weight and a linear layer after layer
normalization. The auxiliary classifier therefore gives probabilities for
the main predictor's weighted objective through a nonlinear projection.

\textbf{Soft-target supervision.}
The auxiliary classifier's detached sigmoid probabilities serve as soft
targets for the main D/A/S head, following the knowledge-distillation
principle~\cite{knowledge_distillation}. For each participant, they
provide task-specific supervision alongside the binary labels. The
supervised and consistency losses, together with their gradient paths,
are defined in Section~\ref{sec:objective}.

\subsection{Prediction and Training Objective}
\label{sec:heads}
\label{sec:objective}
Participant labels supervise predictions from both session and participant
representations. The auxiliary classifier also provides soft targets for
participant predictions. We index participants by $i$ and speaking
conditions by $s$. The prediction head maps the fused participant
representation $\mathbf{u}_{i,\star}$ to three
D/A/S logits. During training, this linear classifier also predicts D/A/S
from each tree-enriched session representation
$\widetilde{\mathbf{h}}_{i,s}$:
\begin{equation}
  \begin{aligned}
    \mathbf{z}_i &=
      \mathbf{W}_{\mathrm{out}}\mathbf{u}_{i,\star}
      +\mathbf{b}_{\mathrm{out}}\in\mathbb{R}^{3},\\
    \mathbf{z}^{\mathrm{sess}}_{i,s} &=
      \mathbf{W}_{\mathrm{out}}\widetilde{\mathbf{h}}_{i,s}
      +\mathbf{b}_{\mathrm{out}}.
  \end{aligned}
  \label{eq:main_logits}
\end{equation}
For tasks $\ell\in\{D,A,S\}$, the sigmoid probabilities are
$p_{i,\ell}=\sigma(z_{i,\ell})$. At test time, we adjust participant logits
with the task-specific offsets described in Section~\ref{sec:bias_adjustment}.

\textbf{Participant and session supervision.}
We train on minibatches of $N$ participants using weighted binary
cross-entropy for the main predictions:
\begin{equation}
  \begin{aligned}
    \mathcal{L}_{\mathrm{part}} = -\frac{1}{3N}
    \sum_{i,\ell}
    \big[&w_\ell y_{i,\ell}\log p_{i,\ell}\\
      &+(1-y_{i,\ell})\log(1-p_{i,\ell})\big].
  \end{aligned}
  \label{eq:participant_loss}
\end{equation}
The positive-class weight is
$w_\ell=\operatorname{clip}(\sqrt{N^-_\ell/N^+_\ell},1,4)$, where
$N^-_\ell$ and $N^+_\ell$ are the negative and positive counts for task
$\ell$ in the official training split. The weights account for differences
in label prevalence across the three targets.

For the session loss $\mathcal{L}_{\mathrm{sess}}$, we apply the same
weighted binary cross-entropy to
$\sigma(\mathbf{z}^{\mathrm{sess}}_{i,s})$. Each valid session uses its
participant's labels, and we average over the three targets and all valid
sessions in the minibatch. This trains the representations of individual
speaking conditions to contribute D/A/S evidence before aggregation.

We also predict the four session types from $\mathbf{h}_{i,s}$, the
encoder output before Session Tree Fusion. A separate classifier uses a
linear layer with 64 hidden units, GELU, and a four-output linear layer.
Its cross-entropy loss $\mathcal{L}_{\mathrm{type}}$ is averaged over valid
sessions and encourages the fused session representation to retain the
session identity supplied by $\mathbf{e}_s$.

\textbf{MLLM supervision and complete objective.}
We compute the auxiliary loss $\mathcal{L}_{\mathrm{M}}$ by applying the
weighted binary cross-entropy in Equation~\ref{eq:participant_loss} to the
auxiliary classifier probabilities $\sigma(\mathbf{z}^{\mathrm{M}}_i)$, using
the same participant labels and class weights. Detaching these
probabilities gives the soft targets
$r_{i,\ell}=\operatorname{stopgrad}(\sigma(z^{\mathrm{M}}_{i,\ell}))$,
which supervise the main classifier through an unweighted consistency loss:
\begin{equation}
  \begin{aligned}
    \mathcal{L}_{\mathrm{cons}} = -\frac{1}{3N}
    \sum_{i,\ell}
    \big[&r_{i,\ell}\log p_{i,\ell}\\
      &+(1-r_{i,\ell})\log(1-p_{i,\ell})\big].
  \end{aligned}
  \label{eq:consistency_loss}
\end{equation}
Stop-gradient directs consistency supervision to the main prediction path.
The auxiliary classifier learns from participant labels on its fixed MLLM
inputs through $\mathcal{L}_{\mathrm{M}}$.

At epoch $e$, the complete objective is
\begin{equation}
  \begin{aligned}
    \mathcal{L}_e ={}& \mathcal{L}_{\mathrm{part}}
      +0.2\mathcal{L}_{\mathrm{sess}}
      +0.15\mathcal{L}_{\mathrm{type}}\\
      &+\eta_e\big(0.05\mathcal{L}_{\mathrm{M}}
      +0.02\mathcal{L}_{\mathrm{cons}}\big),
  \end{aligned}
  \label{eq:total_loss}
\end{equation}
where $\eta_e=\min(e/5,1)$ increases the contribution of both MLLM-related
terms over the first five epochs. The auxiliary classifier is trained
from scratch during main training, so the ramp lets it fit the labels
before its soft targets carry full weight.

\textbf{Input regularization.}
Before encoding during main model training, we add zero-mean Gaussian
noise to valid frames of the audio and video feature sequences. For
participants with at least two available sessions, we also randomly remove
one valid session from the session stream and update the aggregation and
session-supervision mask. The main predictor thus learns from perturbed
features and varying sets of encoded sessions. Noise magnitude,
removal probability, and optimization settings are given in
Section~\ref{sec:optimization}.

%% file: sections/challenge.tex
\section{Task and Evaluation Setting}
\label{sec:challenge}

We evaluate DASWeave on the official AdoDAS Track~1 dataset, following
the prescribed participant split and scoring
protocol~\cite{adodas2026}.

\subsection{Dataset and Released Inputs}
\label{sec:dataset}
The dataset contains 24{,}000 audio-video segments recorded from 6{,}000
children and adolescents at 10 anonymized schools under a standardized
protocol~\cite{adodas2026}. Participants read \emph{The North Wind and the
Sun} in A01. The official training, validation, and test allocations are
shown in Table~\ref{tab:dataset_split}. All four sessions of a participant
belong to the same partition, and every school is represented in all three
partitions.

\begin{table}[H]
  \centering
  \caption{Official participant-disjoint AdoDAS split. Each participant
    contributes four sessions.}
  \label{tab:dataset_split}
  \begin{tabular}{lrr}
    \toprule
    Split & Participants & Segments \\
    \midrule
    Training & 4{,}200 & 16{,}800 \\
    Validation & 600 & 2{,}400 \\
    Test & 1{,}200 & 4{,}800 \\
    \bottomrule
  \end{tabular}
\end{table}

The released inputs comprise feature sequences, pooled acoustic
descriptors, and anonymized ASR transcripts~\cite{adodas2026}. Acoustic and
visual descriptors, pretrained speech and vision embeddings, and voice
activity and video quality signals share a 25~Hz sequence grid. The pooled
descriptors summarize acoustic properties over a session; transcripts are
provided for B01--B03. The feature groups used by
DASWeave are specified in Section~\ref{sec:system}.

\subsection{Track~1 Prediction Targets}
\label{sec:targets}
Participant labels come from the 21-item Depression Anxiety Stress Scales
(DASS-21)~\cite{dass21}, whose item responses lie in $\{0,1,2,3\}$.
Each D/A/S subscale contains seven items. Doubling their sum gives a
score on the conventional 0--42 scale. Track~1 assigns zero to
the \emph{normal} category and one to \emph{mild or above}. A positive label
therefore corresponds to a score of at least 10 for Depression, 8 for
Anxiety, or 15 for Stress.

Positive prevalence in the released training split is 21.55\% for
Depression, 28.24\% for Anxiety, and 12.86\% for Stress. We account for
these class distributions in the weighted training objective in
Section~\ref{sec:objective}.

\subsection{Evaluation Protocol}
\label{sec:evaluation}
The official scorer matches submitted probabilities to anonymized
participant identifiers and uses a fixed decision threshold of 0.5.
Teams are ranked primarily by mean F1 across the three targets:
\begin{equation}
  \begin{aligned}
    \operatorname{F1}_\ell & =
      \frac{2\operatorname{TP}_\ell}
      {2\operatorname{TP}_\ell+\operatorname{FP}_\ell+
       \operatorname{FN}_\ell}, \\
    \operatorname{MeanF1} & =
      \frac{1}{3}\sum_{\ell\in\{D,A,S\}}\operatorname{F1}_\ell,
  \end{aligned}
  \label{eq:challenge_f1}
\end{equation}
where TP, FP, and FN denote true positives, false positives, and false
negatives for each task. F1 is computed for the positive class of each
task. Mean AUROC across D/A/S
breaks ties in the official ranking.

%% file: sections/training.tex
\section{Experimental Setup}
\label{sec:training}
We use the official validation split to select checkpoints and adjust
task-specific biases. The resulting parameters and offsets are then fixed
for hidden-test prediction.

\subsection{Side-Branch Fitting and Representation Extraction}
\label{sec:side_training}
\textbf{Participant folds and OOF extraction.}
The official training participants are divided into five folds, stratified
by their joint D/A/S label pattern. All four sessions of a participant
stay in the same fold. We use seed 2026 for the tree branches and seed 42
for the participant language and MLLM branches. Each training participant's
language and MLLM representations and tree leaves come from side models
fitted on the other four folds; fold masks identify the available tree
inputs. We cache these OOF inputs for main model training, following the
cross-validated construction of second-stage inputs in stacked
generalization~\cite{stacked_generalization}. Median imputers, feature
normalization statistics, and prompt reference quantiles are fitted on
the training participants within each fold.

\textbf{Tree branches.}
We fit both tree branches with a learning rate of 0.03, feature sampling
fraction 0.7, minimum leaf size 20, and $L_2$ regularization coefficient
2.0. Positive-class weights are the negative-to-positive count ratios in
each classifier's training fold. The fixed budget is 800 trees per task
and fold for the session branch and 1,200 for the participant branch,
with at most 31 leaves per tree.

\textbf{Participant language branch.}
For Qwen3, we apply LoRA to the query, key, value, and output projections
with rank 16, $\alpha=32$, and dropout 0.05. Text is limited to 2,048 tokens.
We set the learning rates to $2\times10^{-4}$ for the LoRA parameters and
$10^{-3}$ for the classification head. Training uses one participant per
device with two gradient accumulation steps. Each of the five fold models
is trained for up to five epochs, and its checkpoint is selected by mean
AUROC on the held-out fold.

Validation and test representations come from an additional language
model trained for four epochs on all 4,200 official training participants.
We use its final checkpoint and fit its prompt statistics on the complete
training split.

\textbf{MLLM branch.}
MLLM fitting begins with two models in each participant fold. The
pooled-feature model is trained for up to eight epochs, with checkpoint
selection by held-out mean AUROC. The sequence-feature model uses four
queries per modality and session and is trained for up to twelve epochs;
we select its checkpoint by held-out mean F1. The combined MLLM initializes
its pooled projector, LoRA adapter, and MLLM classification head from the
pooled model and its sequence resampler from the sequence model. It keeps
the first two of the sequence model's four learned queries, giving two
tokens per modality and session for the prefix described in
Section~\ref{sec:mllm}.

We train the combined MLLM for up to eight epochs using the participant
language branch's LoRA configuration. During the first epoch, the adapter
is frozen while the projectors and head are trained; subsequent updates
are joint. Learning rates are $2\times10^{-5}$ for LoRA,
$5\times10^{-5}$ for the pooled projector, $10^{-4}$ for the sequence
projector, and $2\times10^{-4}$ for the head. The text limit is 512 tokens;
training uses one participant per device and eight gradient accumulation
steps. After selecting the combined checkpoint by held-out mean AUROC, we
extract OOF representations for auxiliary supervision during main training.
Appendix~\ref{sec:appendix} gives the remaining projector settings and prompt
structures.

\subsection{Main Model Optimization}
\label{sec:optimization}
The main model is optimized with AdamW~\cite{adamw}. Learned tree modules
and Participant Language Fusion form a separate parameter group with a
lower learning rate. We keep MacBERT frozen and the prepared side inputs
fixed, and initialize the biases of the main D/A/S head and auxiliary
classifier to the log-odds of the training label prevalence. Linear
learning-rate warmup is followed by cosine decay.
Table~\ref{tab:training_config} in Appendix~\ref{sec:appendix_inputs} lists
the optimization settings and the noise magnitude and session removal
probability used for input regularization.

\subsection{Checkpoint Selection and Averaging}
\label{sec:averaging}
After each epoch, we evaluate participant predictions on the official
validation split. Early stopping monitors weighted participant validation
loss with a patience of eight epochs. To rank checkpoints, we instead use
mean F1 after the task-specific bias adjustment in
Section~\ref{sec:bias_adjustment}, retaining the three highest-scoring
snapshots from the run.

We then average the parameters of the three selected checkpoints
uniformly, following parameter-space averaging~\cite{weight_averaging}:
\begin{equation}
  \overline{\boldsymbol{\theta}} =
  \frac{1}{3}\sum_{k=1}^{3}\boldsymbol{\theta}_k.
  \label{eq:checkpoint_average}
\end{equation}
This gives a single prediction model. We collect its validation logits
and repeat the task-specific offset search, so that the final offsets
correspond to the averaged model used for test prediction.

\subsection{Task-Specific Bias Adjustment}
\label{sec:bias_adjustment}
Track~1 computes each target's F1 at a fixed probability threshold of 0.5.
An additive logit offset selected for each task sets its
precision--recall operating point~\cite{f1_thresholding}. For candidate
offset $\delta$, the validation decision is
$\widehat{y}_{i,\ell}(\delta)=
\mathbb{I}[\sigma(z_{i,\ell}+\delta)>0.5]$. We search
$\mathcal{B}=\{-3.0,-2.9,\ldots,3.0\}$ and choose
\begin{equation}
  \delta^{\star}_\ell =
  \underset{\delta\in\mathcal{B}}{\operatorname{arg\,max}}
  \operatorname{F1}^{\mathrm{val}}_\ell(\delta).
  \label{eq:bias_selection}
\end{equation}
We resolve ties by taking the first maximum in ascending grid order.
All tasks and ablation configurations share the range, step size, and
selection criterion. Since each offset affects only its own task's
decisions, the three independent searches maximize mean F1 over the joint
grid.

The sigmoid is monotone, so the adjusted rule is equivalent to
$z_{i,\ell}>-\delta_\ell$, or to applying a threshold of
$\sigma(-\delta_\ell)$ to the unshifted probability. This constant shift
sets the operating point used by the official scorer and preserves the
within-task ranking. We submit the probabilities
\begin{equation}
  p^{\mathrm{sub}}_{i,\ell} =
  \sigma\big(z_{i,\ell}+\delta^{\star}_\ell\big).
  \label{eq:submission_probability}
\end{equation}
The averaged model's selected D/A/S offsets are $(-0.6,0,0)$. We fix them
for the test logits and return one probability triplet per participant.

%% file: sections/results.tex
\section{Experimental Results}
\label{sec:results}
Official Track~1 hidden-test results assess the complete method. We then
examine its representation and auxiliary supervision components through
validation ablations.

\subsection{Official Track~1 Results}
\label{sec:official_results}
\label{sec:per_task_results}


\begin{table}[H]
  \centering
  \caption{Official Track~1 hidden-test results of the organizer's audio-visual
    baseline and the top five teams~\cite{adodas2026}. Our submission is
    highlighted in bold.}
  \label{tab:official_leaderboard}
  \small
  \setlength{\tabcolsep}{3pt}
  \begin{tabular*}{\columnwidth}{@{\extracolsep{\fill}}rlrr@{}}
    \toprule
    Rank & Method & Mean F1 $\uparrow$ & Mean AUROC $\uparrow$ \\
    \midrule
    -- & Organizer A+V & 0.4604 & 0.7169 \\
    \midrule
    1 & HITSZ\_HLT & 0.5921 & 0.7652 \\
    \textbf{2} & \textbf{USTC-AC (ours)} & \textbf{0.5853} & \textbf{0.8019} \\
    3 & cog-simulation & 0.4747 & 0.7051 \\
    4 & HEALTH FIRST & 0.4710 & 0.7320 \\
    5 & Ritsumei & 0.4665 & 0.7282 \\
    \bottomrule
  \end{tabular*}
\end{table}

\begin{table}[t]
  \centering
  \caption{Official hidden-test scores of the USTC-AC submission by task.}
  \label{tab:official_per_task}
  \begin{tabular}{@{}lrr@{}}
    \toprule
    Target & F1 $\uparrow$ & AUROC $\uparrow$ \\
    \midrule
    Depression & 0.6288 & 0.8134 \\
    Anxiety & 0.6421 & 0.7962 \\
    Stress & 0.4851 & 0.7960 \\
    \midrule
    Mean & \textbf{0.5853} & \textbf{0.8019} \\
    \bottomrule
  \end{tabular}
\end{table}

\begin{table*}[!t]
  \centering
  \caption{Ablation results on the official validation split (600
    participants, seed 123). Each configuration selects task-specific offsets
    using the same procedure. $\Delta$ mean F1 is the difference
    from the full method, calculated before rounding. Best scores are bold.}
  \label{tab:validation_ablation}
  \begin{tabular*}{\textwidth}{@{\extracolsep{\fill}}lrrrrr@{}}
    \toprule
    Configuration & F1-D $\uparrow$ & F1-A $\uparrow$ & F1-S $\uparrow$ & Mean F1 $\uparrow$ & $\Delta$ mean F1 $\uparrow$ \\
    \midrule
    Backbone (audio-visual + MacBERT) & 0.5704 & 0.6007 & 0.4205 & 0.5305 & $-0.0733$ \\
    \textbf{Full method} & \textbf{0.6377} & \textbf{0.6520} & \textbf{0.5217} & \textbf{0.6038} & $0$ \\
    w/o participant language fusion & 0.5904 & 0.6209 & 0.4699 & 0.5604 & $-0.0434$ \\
    w/o two-level tree fusion & 0.6103 & 0.6354 & 0.4809 & 0.5755 & $-0.0283$ \\
    w/o MLLM auxiliary supervision & 0.6277 & 0.6449 & 0.4973 & 0.5900 & $-0.0138$ \\
    w/o MacBERT & 0.6327 & 0.6447 & 0.5054 & 0.5943 & $-0.0095$ \\
    \bottomrule
  \end{tabular*}
\end{table*}

Submitted under the team name USTC-AC, DASWeave ranked second among 95
eligible Track~1 teams on the official hidden test set, with mean F1 of
0.5853 and mean AUROC of 0.8019~\cite{adodas2026}.

Table~\ref{tab:official_leaderboard} compares DASWeave with the organizer's
audio-visual baseline. Mean F1 increases by 0.1249
(27.1\% relative) and mean AUROC by 0.0850. In mean F1, DASWeave is 0.0068
below the first-place submission and 0.1106 above the third-place
submission. It also has the highest mean AUROC among the top five Track~1
teams, 0.0367 above the first-place
submission. Across the three targets in Table~\ref{tab:official_per_task},
AUROC is at least 0.7960.

\subsection{Ablation Study}
\label{sec:ablation_protocol}
\label{sec:component_results}
\textbf{Configurations and matched protocol.}
The comparison on the official validation split of 600 participants
includes six configurations. The backbone keeps only the Shared Session
Encoder, with its audio-visual and local text features, followed by
participant aggregation and the prediction head. The full method adds
Two-Level Tree Fusion, Participant Language Fusion, and MLLM Auxiliary
Supervision. Each of the other four configurations removes one of these
three components or the local text features from the full method and keeps
everything else. The tree ablation removes both session and participant
tree fusion, and the MLLM ablation removes the auxiliary classification and
consistency losses together.

We retrain the backbone and each removal configuration on the
official training split, reusing fixed OOF inputs for the side branches
that remain. All six configurations share seed 123, the training budget,
checkpoint-selection rules, and parameter-averaging procedure described in
Section~\ref{sec:training}. Each selects its own task-specific offsets
from its validation logits using the common search procedure in
Section~\ref{sec:bias_adjustment}. The resulting F1 scores, after bias
adjustment and at the challenge's fixed threshold of 0.5, are reported in
Table~\ref{tab:validation_ablation}.

\textbf{Complete method.}
With mean F1 of 0.6038, DASWeave improves on the backbone's 0.5305 by
7.33 percentage points and has the highest F1 on all three targets among
the six configurations. Each of the three components and the local text
features improves F1 on all three targets.

\textbf{Participant Language Fusion.}
Without Participant Language Fusion, mean F1 falls to 0.5604, the largest
decrease at 4.34 percentage points. This component is the only part of the
prediction path that encodes all answers together, and it also describes
how speaking behaviour differs from the reading. It therefore adds
information beyond the temporal features, tree tokens, and local text
features.

\textbf{Two-Level Tree Fusion.}
Without the two tree fusion branches, mean F1 is 0.5755, 2.83 percentage
points lower. The tree tokens bring behavioural statistics into the
predictor, including differences across speaking conditions, and they add
information even with Participant Language Fusion in place.

\textbf{MLLM Auxiliary Supervision.}
Training without the auxiliary classification and consistency losses gives
mean F1 of 0.5900, 1.38 percentage points lower. This configuration and
the full method use the same prediction path formed by components A--C, so
the gain comes from training with joint audio-visual-text supervision.

\textbf{Local text features.}
Without the MacBERT text features, mean F1 falls by 0.95 percentage points
to 0.5943, although Participant Language Fusion still encodes all answers.
The local text features of each answer therefore complement the language
representation of all answers.

\subsection{Analysis and Evaluation Scope}
\label{sec:analysis_scope}
\label{sec:evaluation_scope}
\textbf{Evidence across speaking conditions.}
Participant Language Fusion and Two-Level Tree Fusion bring evidence across
speaking conditions into the prediction path, and they give the two largest
gains. This result supports the motivation in
Section~\ref{sec:introduction} that comparing a participant's responses
adds information beyond individual responses.

\textbf{Cross-task behaviour.}
For Stress, official AUROC is close to that for Anxiety, although its F1
is lower, as shown in Table~\ref{tab:official_per_task}. AUROC evaluates
participant ordering across thresholds, whereas F1 evaluates decisions at
the selected operating point. Stress also has the lowest positive
prevalence, 12.86\% in the training split, so its F1 depends on fewer
positive participants.

Stress has the largest validation improvement over the backbone, from
F1 of 0.4205 to 0.5217, a gain of 10.12 percentage points.
Table~\ref{tab:validation_ablation} also shows that Stress has the largest
F1 decrease after each of the four removals. All three components and the
local text features therefore contribute most to the least prevalent
target.

\textbf{Evaluation setting.}
All ten schools appear in each official partition. The results therefore
measure DASS-21-based screening of unseen participants under broadly
matched collection conditions.

%% file: sections/conclusion.tex
\section{Conclusion}
\label{sec:conclusion}

DASWeave brings behavioural differences across speaking conditions into
multimodal D/A/S screening. Two-Level Tree Fusion adds tree tokens of
session statistics before session aggregation and of participant
statistics after it. The participant statistics include differences
between speaking conditions. Participant Language Fusion encodes all
answers together with descriptions of how speaking behaviour differs from
the reading. During training, MLLM Auxiliary Supervision brings joint
audio-visual-text information into the main predictor through soft
targets.

In the matched validation ablations, each of the three components and the
local text features improves F1 on all three targets. Participant Language
Fusion and Two-Level Tree Fusion, which bring evidence across speaking
conditions into the prediction path, contribute the most, and Stress has
the largest gain over the backbone. DASWeave ranked second on AdoDAS
Track~1 and achieved the highest mean AUROC among the top five teams.

The gains for Stress motivate closer study of task-specific supervision
and cross-session aggregation. Evaluating transfer to held-out schools and
different recording conditions is another direction for future work.

%% file: sections/appendix.tex
\section{Implementation Details}
\label{sec:appendix}

\subsection{Inputs and Branch Configuration}
\label{sec:appendix_inputs}
Tables~\ref{tab:training_config} and~\ref{tab:branch_configuration}
summarize the main training configuration and branch architectures.
We order every participant's inputs by session A01, B01, B02, B03 and
target D, A, S. The main encoder matches each sequence to a 40~ms grid
using the nearest timestamp within a tolerance of 25~ms. For audio,
validity is the intersection of the available log-Mel/MFCC and
speech-embedding masks; for video, it is the intersection of the available
visual-embedding and quality masks. If a modality has no available core
group, we use the union of its other group masks. Missing feature groups
and pooled acoustic descriptors are filled with zeros. An absent session
likewise receives zero-filled inputs and a false session-validity mask.
We collate reading and open-response sessions separately, padding
variable-length sequences to the longest in each group.

MacBERT takes a Chinese question--response string limited to 512 tokens.
We concatenate normalized ASR segments in their stored order to form the
answer, keeping the question when the answer is empty. A01 has a zero text
representation. We retain the extracted scale of the participant Qwen3
representation and $L_2$-normalize the MLLM representation before the
auxiliary classifier. For $N$ participants, the session and participant
tree leaf arrays have shapes
$N\times4\times3\times5\times800$ and $N\times3\times5\times1200$
respectively. The axes denote participants, sessions (session trees only),
targets, folds, and trees.

\begin{table}[H]
  \centering
  \caption{Main model training configuration. Side-branch fitting settings
    are given in Section~\ref{sec:side_training}.}
  \label{tab:training_config}
  \small
  \begin{tabular}{@{}p{0.58\columnwidth}p{0.35\columnwidth}@{}}
    \toprule
    Setting & Value \\
    \midrule
    Main learning rate & $10^{-3}$ \\
    Tree / language fusion learning rate & $10^{-4}$ \\
    Weight decay & 0.01 \\
    Participant batch size & 48 \\
    Maximum epochs & 30 \\
    Learning-rate warmup & 3 epochs \\
    Early-stopping patience & 8 epochs \\
    Dropout & 0.2 \\
    Feature noise standard deviation & 0.01 \\
    Session removal probability & 0.1 \\
    Gradient clipping norm & 1.0 \\
    Mixed precision & BF16 \\
    Random seed & 123 \\
    \bottomrule
  \end{tabular}
\end{table}

\begin{table*}[t]
  \centering
  \caption{Branch inputs and architecture settings. The tree dimensions
    include all selected descriptors and session indicators where applicable.
    Side-branch fitting and main optimization settings are given in
    Sections~\ref{sec:side_training} and~\ref{sec:optimization}.}
  \label{tab:branch_configuration}
  \small
  \setlength{\tabcolsep}{4pt}
  \begin{tabular}{@{}>{\raggedright\arraybackslash}p{0.16\textwidth}
      >{\raggedright\arraybackslash}p{0.39\textwidth}
      >{\raggedright\arraybackslash}p{0.39\textwidth}@{}}
    \toprule
    Component & Inputs & Architecture \\
    \midrule
    Shared Session Encoder &
      Audio: log-Mel/MFCC, VAD, Chinese XLSR wav2vec~2.0, pooled eGeMAPS.
      Video: head pose, facial behaviour, quality, aggregated VAD,
      SigLIP-SO400M. Text: MacBERT-base. &
      Feature adapters: 64; TCN width: 256; 6 blocks, kernel 3;
      pooling: 512 per modality. Text: $768\rightarrow64$;
      pooled acoustic projection: 64; session embedding: 16;
      session output: 256. \\
    Session tree &
      HuBERT-large and ViT-base pooled means/stds, low-dimensional
      descriptors, and four session indicators: 4{,}207 features. &
      5 folds; 800 trees per task; at most 31 leaves per tree.
      Leaf vocabulary: 32; embedding width: 32;
      output: three 256-dimensional tree tokens. \\
    Participant tree &
      HuBERT-large and DINOv2-large pooled means/stds, low-dimensional
      descriptors, and cross-session statistics: 47{,}150 features. &
      5 folds; 1{,}200 trees per task; at most 31 leaves per tree.
      Leaf vocabulary: 32; embedding width: 32;
      output: three 256-dimensional tree tokens. \\
    Participant language &
      Qwen3-8B; structured four-session prompt;
      text limit: 2{,}048 tokens. &
      Final-token state: 4{,}096; fusion projection:
      $4{,}096\rightarrow512\rightarrow256$;
      4 attention heads; one key/value token. \\
    MLLM Auxiliary Supervision &
      Qwen3-8B; HuBERT-large and DINOv2-base.
      Sequence widths: 1{,}024/768; pooled widths: 5{,}120/3{,}840
      (audio/video). Text limit: 512 tokens. &
      Prefix: 8 pooled + 16 sequence tokens; 2 queries per modality/session.
      Final-token state: 4{,}096;
      main-model auxiliary projection: 256. \\
    \bottomrule
  \end{tabular}
\end{table*}

The MLLM resampler uses width 512, two blocks, eight attention heads, and
dropout 0.1. For the pooled projectors, hidden width is 1{,}024 and dropout
is 0.1. We initialize the combined MLLM's pooled and sequence prefix scales
to 0.08 and 0.01. During training, each complete prefix block is
independently removed per participant, with probability 0.05 for the pooled
block and 0.2 for the sequence block. Input scalers estimate per-dimension
means and standard deviations from valid features in the training folds
specified in Section~\ref{sec:side_training}.

\subsection{Prompt Construction}
\label{sec:appendix_prompts}
The two language branches use Chinese prompts. Below, we give their
structure in English. Transcripts, emotion tags, measured
descriptors, and training-derived reference tags fill the dynamic fields.
We wrap the participant language prompt in the Qwen3 chat template,
disable thinking, and append a Chinese summary anchor to the assistant
prefix. Both Qwen3 branches use left padding and left truncation. Each
extracts the hidden state of the final text token.

\begin{quote}
  \small\raggedright
  \setlength{\parskip}{0pt}
  \textbf{Participant language prompt.}\par
  Task: encode evidence for D/A/S classification over the past week.\par
  Task cues: low mood, motivation and interest for D; fear, worry and
  tension for A; academic demands, sleep, irritability, relaxation and
  interpersonal conflict for S.\par
  A01: reading baseline; acoustic and visual/motion summaries.\par
  B01--B03, in order: question, transcript, SenseVoice emotion summary,
  acoustic summary, visual/motion summary, and response observations when
  applicable.\par
  Cross-session overview: response duration and text length, short-response
  counts, loudness and pitch-range ratios to reading, and visual quality.\par
  Final request: form the D/A/S representation; append the summary anchor.
\end{quote}

Acoustic summaries contain duration, speech ratio, pauses, pitch,
loudness, and harmonic-to-noise ratio. The reading summary also gives
speaking rate. Visual summaries cover detection and validity rates,
expression changes, head movement, global motion, and body-keypoint
variation. To assign relative position tags, we use the training-fold
10th, 25th, 75th, and 90th percentiles alongside fixed descriptive anchors.
Missing numeric values appear as \texttt{NA}.

\begin{quote}
  \small\raggedright
  \setlength{\parskip}{0pt}
  \textbf{MLLM text prompt.}\par
  Task: assess D/A/S from the preceding audio-visual representation tokens
  and the following response text.\par
  Prefix description: pooled representations and supplementary sequence
  representations from all four sessions and both modalities.\par
  B01--B03, in order: question, transcript, SenseVoice emotion tags.\par
  Final request: provide internal representations for the three binary
  targets D, A, and S.
\end{quote}

In the MLLM input, pooled tokens come before sequence tokens. Within each
block, audio precedes video, and each modality follows the four-session
order. Sequence queries are grouped by session. We apply the 512-token
text limit before prepending the 24 feature tokens.

\newpage
\subsection{Reproduction Workflow and Dependencies}
\label{sec:appendix_reproduction}
Table~\ref{tab:branch_usage} distinguishes parameters learned during main
training from prepared side representations. Reproduction follows these
steps:
\begin{enumerate}
  \setlength{\itemsep}{2pt}
  \setlength{\parsep}{0pt}
  \item Align features and transcripts by participant and session while
    preserving the official participant split.
  \item Fit side models in five participant folds. Fit imputation,
    feature scalers, and prompt statistics within each fold, then extract
    each training participant's side inputs from models fitted on the
    other four folds.
  \item For validation and test, prepare leaves from all five tree models
    and participant language representations from the Qwen3 model fitted
    on the complete training split. Cache
    the resulting prediction inputs with their participant identifiers.
  \item With the side inputs fixed, train the participant model using the
    objective and input regularization in Section~\ref{sec:objective}
    and the optimization settings in
    Section~\ref{sec:optimization}.
  \item Select the three checkpoints with the highest bias-adjusted
    validation mean F1 and average them. Use the averaged model's
    validation logits to select offsets.
    The submission uses D/A/S offsets $(-0.6,0,0)$.
  \item Run the averaged main model on test inputs with the fixed offsets
    to return one D/A/S probability triplet per participant.
\end{enumerate}

\begin{table}[H]
  \centering
  \caption{Component use during main model training and prediction.
    Side-model fitting precedes main model training.}
  \label{tab:branch_usage}
  \small
  \setlength{\tabcolsep}{3pt}
  \begin{tabular}{@{}>{\raggedright\arraybackslash}p{0.37\columnwidth}
      >{\raggedright\arraybackslash}p{0.28\columnwidth}
      >{\raggedright\arraybackslash}p{0.27\columnwidth}@{}}
    \toprule
    Component & Main training & Prediction \\
    \midrule
    Shared Session Encoder & Optimize & Forward \\
    MacBERT encoder & Frozen & Forward \\
    Tree classifiers & Fixed OOF leaves & All-fold leaves \\
    Tree leaf embeddings and fusion & Optimize & Forward \\
    Language projection and fusion & Optimize & Forward \\
    Participant language branch & Fixed OOF states & Full-training model \\
    MLLM & Fixed OOF states & -- \\
    Auxiliary classifier & Optimize & -- \\
    \bottomrule
  \end{tabular}
\end{table}

\begin{samepage}
For main inference, the submitted environment specifies Python 3.10,
PyTorch 2.5.1, and Transformers 4.46.3. The Qwen3 branches have separate
environments specifying Transformers 4.51.3, PEFT 0.15.2, and Accelerate
1.6.0. The specified LightGBM version range for the tree branch is
$3.3.5\leq v<5$. Separate environments serve main prediction and
side-representation preparation.\par
\end{samepage}